\documentclass[
twocolumn]{revtex4-1}

\usepackage{graphicx}
\usepackage{epstopdf}
\usepackage{dcolumn}
\usepackage{bm}
\usepackage{amsmath}
\usepackage{amssymb}
\usepackage{mathrsfs}
\usepackage{accents}
\usepackage{cancel}
\usepackage{hyperref}

\begin{document}


\title{Explanation of Gravity Probe B Experimental Results using Heaviside-Maxwellian (Vector) Gravity in flat Space-time}

\author{Harihar Behera}
\email{behera.hh@gmail.com}
\affiliation{
BIET Degree College and BIET Higher Secondary School, Govindpur, Dhenkanal-759001, Odisha, India.}
\author{Niranjan Barik}
\email{dr.nbarik@gmail.com}
\affiliation{Department of Physics, Utkal University, Vani Vihar, Bhubaneswar-751004, Odisha, India}




\date{\today}

\begin{abstract}
The Gravity Probe B (GP-B) Experiment of NASA was aimed to test the theoretical predictions of Einstein's 1916 relativistic tensor theory of gravity in curved space-time (General Relativity (GR)) concerning the spin axis precession of a gyroscope moving in the field of a slowly rotating massive body, like the Earth.  In 2011, GP-B mission reported its measured data on the precession (displacement) angles of the spin axes of the four spherical gyroscopes housed in a satellite orbiting 642 km (400 mi) above the Earth in polar orbit. The reported results are in agreement with the predictions of GR. For the first time, here we report an undergraduate level explanation of the GP-B experimental results using Heaviside-Maxwellian (vector) Gravity (HMG) in flat space-time, first formulated by Heaviside in 1893, and later considered/re-discovered by many authors. Our new explanation of the GP-B results provides a new test of HMG apart from the existing ones, which deserves the attention of researchers in the field for its simplicity and new perspective. 
 
\end{abstract}

\maketitle
\section{Introduction}
\label{intro}
Gravity Probe B (GP-B), launched on 20 April 2004, was a NASA physics mission to experimentally investigate Einstein's 1916 theory of general relativity (GR) - his relativistic theory of gravity in curved space-time. GP-B used four spherical gyroscopes and a telescope, housed in a satellite orbiting 642 km (400 mi) above the Earth in a polar orbit, to measure two of its significant predictions (with unprecedented accuracy) on the precession of the spin axis of a gyroscope moving in the field of the Earth predicted by Pugh \cite{1} in 1959 and by Schiff \cite{2,3,4}, in 1960, using GR. This is called Schiff effect/precession, which was measured in a satellite (a co-moving frame) in which gyroscopes are at rest. In the context of GR, the working formula for the GP-B experiment in the polar orbit (Schematic in figure 1) was worked out by Schiff \cite{2,3,4} as 
\begin{equation}
\frac{d\mathbf{s}}{d\tau} = \left(\mathbf{\Omega}_{gd}^{\text{GR}} + \mathbf{\Omega}_{fd}^{\text{GR}}\right)\times \mathbf{s} \label{eq1}
\end{equation}
where $\mathbf{s}$ is the spin angular momentum vector of a gyroscope in its rest frame, $\tau$ is the time measured in gyro's rest frame (i.e. the satellite), called proper time, vectors $\mathbf{\Omega}_{gd}^{\text{GR}}$ representing the Geodetic Effect and $\mathbf{\Omega}_{fd}^{\text{GR}}$ Frame-dragging effect, as they are called in the GR parlance, are given by
\begin{eqnarray}
\mathbf{\Omega}_{gd}^{\text{GR}} = \frac{3GM}{2c^2r^3}\left(\mathbf{r}\times \mathbf{v}\right)    \label{eq2}  \\
\mathbf{\Omega}_{fd}^{\text{GR}} = \frac{G}{c^2r^3}\left[\frac{3\mathbf{r}}{r^2}(\mathbf{r}\cdot \mathbf{S}_E) - \mathbf{S}_E\right]  \label{eq3}
\end{eqnarray}
$G$ being the gravitational constant, $c$ is the velocity of light in vacuum, $\mathbf{S}_E = I\mathbf{\omega}_E$ is the spin angular momentum of the Earth ($I$ and $\mathbf{\omega}_E$ are Earth's moment of inertia and angular velocity  respectively), $\mathbf{r}$ is the position vector of a gyroscope from Earth's center and $\mathbf{v}$ is the orbital velocity of a gyroscope. The predicted quantities $\mathbf{\Omega}_{gd}^{\text{GR}}$ in eq. \eqref{eq2} and  $\mathbf{\Omega}_{fd}^{\text{GR}}$ in eq. \eqref{eq3} could be computed since each quantity in the equation eqs. \eqref{eq2}-\eqref{eq3} was known: the instantaneous orbital velocity and radius from GP-B’s on-board GPS detector; the mass, moment of inertia, and angular velocity of the Earth from geophysical and astrophysical data. As we know, GP-B experiment has confirmed both significant predictions of GR: the geodetic effect \eqref{eq2} to a precision of $0.3\%$ and frame-dragging \eqref{eq3} to $20\%$ \cite{5,6}. However, in this communication we show how one can obtain the eqs. \eqref{eq1}-\eqref{eq3} within the framework of Heaviside-Maxwellian Gravity (HMG) in flat space-time briefly described below.  
\begin{figure}
\includegraphics[scale=1.35]{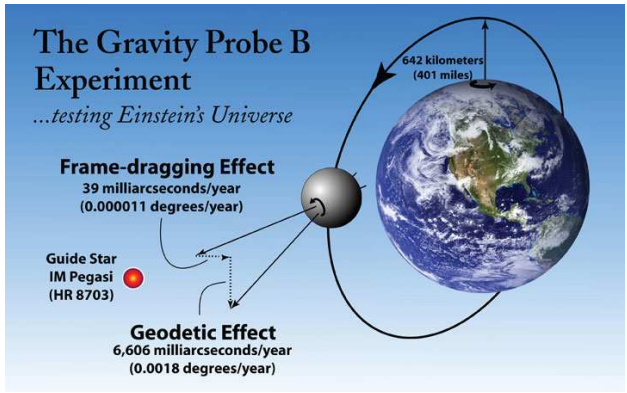}
\caption{Schematic of the orbit of the Gravity Probe
B satellite. The Geodetic and Frame-dragging effects. North–South, East–West relativistic precessions with respect to the guide star IM Pegasi for an ideal gyroscope in polar orbit around the Earth. [Figure Courtesy: \href{https://einstein.stanford.edu/Media/Everitt_Lecture-051806.pdf}{C. W. F. Everitt.}}
\end{figure}

\section{Heaviside-Maxwellian Gravity (HMG)} 
The fundamental equations of HMG are analogous to the Maxwell-Lorentz equations of electromagnetism and are called gravito-Maxwell-Lorentz equations (g-MLEs). Recently Behera \cite{7}, using Galileo Newtonian Relativity (GNR) and Behera and Barik \cite{8}, using the U(1) local phage (or gauge) in-variance of Dirac's massive field theory (of charged as well as neutral particles), have found two equivalent mathematical representations of HMG: (1) Heaviside Gravity (HG) as originally written down by Heaviside in 1893 \cite{9,10,11} and (2) Maxwellian Gravity (MG) \cite{7,8,12,13}. The g-MLEs of HG and MG, which represent the same physical theory HGM, are listed in Table 1. It is to be noted that, Heaviside speculated a gravito-Lorentz force with a wrong sign in the velocity dependent term by electromagnetic analogy, i.e. of MG-type in Table 1. But the correct form is given in Table 1, as derived Behera \cite{7} using Schwinger's formalism within GNR and  Behera and Barik \cite{8} using the principle of local phase (or gauge) invariance of Dirac's massive quantum field theory .  \\ 
\begin{table}[h]
\begin{center}
   \begin{tabular}{ | l | l |}
    \hline
    gravito-MLEs of MG & gravito MLEs of HG \\ \hline
    $\mathbf{\nabla}\cdot\mathbf{g} = - 4\pi G \rho_g = - \frac{\rho_g}{\epsilon_{0g}}$ & $\mathbf{\nabla}\cdot\mathbf{g} = - 4\pi G \rho_g = - \frac{\rho_g}{\epsilon_{0g}}$ \\ \hline
    $\mathbf{\nabla}\cdot\mathbf{b} = 0$ & $\mathbf{\nabla}\cdot\mathbf{b} = 0$ \\ \hline
    $\mathbf{\nabla}\times\mathbf{b}\,=\,-\,\mu_{0g}\mathbf{j}_g\,+\,\frac{1}{c_g^2}\frac{\partial \mathbf{g}}{\partial t}$ & $\mathbf{\nabla}\times\mathbf{b}\,=\,+\,\mu_{0g}\mathbf{j}_g\,-\,\frac{1}{c_g^2}\frac{\partial \mathbf{g}}{\partial t}$  \\ \hline 
    $\mathbf{\nabla}\times\mathbf{g}\,=\,-\,\frac{\partial \mathbf{b}}{\partial t}$ & $\mathbf{\nabla}\times\mathbf{g}\,=\,+\,\frac{\partial \mathbf{b}}{\partial t}$  \\ \hline
    $\frac{d\mathbf{p}}{dt}\,=\,m_g\left[\mathbf{g}\,+\,\mathbf{u}\times \mathbf{b}\right]$ & $\frac{d\mathbf{p}}{dt}\,=\,m_g\left[\mathbf{g}\,-\,\mathbf{u}\times \mathbf{b}\right]$  \\ \hline 
   $\mathbf{b}\,=\,+\mathbf{\nabla}\times \mathbf{A}_g$ & $\mathbf{b}\,=\,-\mathbf{\nabla}\times \mathbf{A}_g$ \\ \hline
     $\mathbf{g}\,=\,-\,\mathbf{\nabla}\phi_{g} \,-\,\frac{\partial \mathbf{A}_g}{\partial t}$ & $\mathbf{g}\,=\,-\,\mathbf{\nabla}\phi_{g} \,-\,\frac{\partial \mathbf{A}_g}{\partial t}$ \\ \hline
\end{tabular}
\end{center}
\caption{Physically Equivalent Sets of gravito-Maxwell-Lorentz Equations (g-MLEs) representing Heaviside-Maxwellian Gravity (HMG).}
    \label{tab:truthTables}   
\end{table}
In Table 1, in analogy with Maxwell-Lorentz equations of classical electromagnetism in SI units, we have introduced two new universal constants for vacuum: 
\begin{equation}
\epsilon_{0g} = \frac{1}{4\pi G}, \quad \mu_{0g} = \frac{4\pi G}{c_g^2} \quad \Rightarrow \quad c_g = \frac{1}{\sqrt{\epsilon_{0g}\mu_{0g}}}  \label{eq4}
\end{equation}
where the speed of gravitational waves in vacuum $c_g = c$, as theoretically shown in ref. \cite{8,12,13} - an important prediction of HMG which agree well with the recent (almost) simultaneous detection of the Gravitational Waves and Gamma-Rays from a Binary Neutron Star Merger events: GW170817 and GRB 170817A \cite{14}, gravitatinal mass density $\rho_g = \rho_0$, the positive (rest) mass density (as theoretically proved in \cite{8,12,13} without sacrificing the time-tested empirical law of universality of free fall of Galileo), $\mathbf{j}_g = \rho_0\mathbf{v}$ stands for the mass current density ($\mathbf{v}$ is velocity of $\rho_0$) and by electromagnetic analogy, $\mathbf{b}$ may be called the gravito-magnetic field, the Newtonian gravitational field $\mathbf{g}$ may be called as gravito-electric  field, $\epsilon_{0g}$ and $\mu_{0g}$ may be named respectively as the gravito-electric (or gravitic) permitivity and the gravito-magnetic permeability of vacuum. 
The g-MLEs of MG-type have also been derived from other approaches to gravito-electromagnetic (GEM) theory using (a) Galileo-Newtonian Relativity \cite{13}, (b) special relativity  \cite{12,13,15,16}, (c) principle of causality \cite{17,18} (d) common axiomatic methods applicabe to electricity and gravity \cite{19,20} and (e) a specific approach to linearize  
GR in weak field and slow motion approximation \cite{21}. These variant of approaches that lead to HMG in its either form, establish g-MLEs as self-consistent. It is to be noted that the explanations for the (a) perihelion advance of Mercuty (b) gravitational bending of light and (c) the Shapiro time delay within the vector theory of gravity exist in the literature \cite{22,23,24,25}. Recently Hilborn \cite{26} following an electromagnetic analogy, calculated the wave forms of gravitational radiation emitted by orbiting binary objects that are very similar to those observed by the Laser Interferometer Gravitational-Wave Observatory (LIGO-VIRGO) gravitational wave collaboration in 2015 up to the point at which the binary merger occurs. Hilborn's calculation produces results that have the same dependence on the masses of the orbiting objects, the orbital frequency, and the mass separation as do the results from the linear version of general relativity (GR). But the polarization, angular distributions, and overall power results of Hilborn differ from those of GR. Recognizing this and out of scientific curiosity, we ventured to adopt the MG form of HMG in Table 1 to calculate the spin axis precession in the gravito-electromagnetic (GEM) field of the spinning Earth according to MG in flat space-time by electromagnetic analogy as under.         
\section{Spin Gravitomagnetic Moment and associated Gravitomagenic Field:} 
For the purpose of the present paper, the relevant electromagnetic analogy is the motion of a small spherical uniformly charged sphere (corresponding to a gyroscope in GP-B Experiment) in an orbit around another huge spherical spinning charged sphere (coresponding to the huge Earth in GP-B Experiment). In electromagnetism, a spinning charged sphere creates a di-polar magnetic field $\mathbf{B}$ around it and this field is related to its spin magnetic moment which is proportional to the spin angular momentum of the spinning object. Kramers \cite{27}, Corben \cite{28} and Bohm \cite{29} have provided us classical derivations of spin magnetic moment of a spinning charged particle, which precisely matches with Dirac's quantum mechanical finding the of spin magnetic moment of the electron:
\begin{equation}
\mathbf{\mu}_s = (q/m_0) \mathbf{s} = (g_sq/2m_0) \quad \text{(in SI units)},  \label{eq5} 
\end{equation}
where $q = - |e|$ is the charge and $\mathbf{s}$ is the spin angular momentum the electron, $g_s$ is the so-called $g$-factor associated with the spin magnetic moment of a charged particle - called the gyromagnetic ratio; $g_s = 2$ for an electron as found by Dirac quantum mechanically and by others \cite{27,28,29} classicaly. In classical electromagnetism, equation \eqref{eq5} also holds good for a uniformly magnetized sphere \cite{30}. The spin magnetic moment produces a di-polar magnetic field $\mathbf{B}(\mathbf{r})$ at a field point $\mathbf{r} \neq \mathbf{0}$ in vacuum \cite{30}: 
\begin{equation} 
\mathbf{B}(\mathbf{r}) = \frac{\mu_0}{4\pi}\left[\frac{3\mathbf{\hat{r}}(\mathbf{\hat{r} }\cdot \mathbf{\mu}_s) - \mathbf{\mu}_s }{r^3} \right] \quad \text{(in SI units)}, \label{eq6}  
\end{equation}
where $\mu_0$ is the magnetic permeability of vacuum and $\mathbf{\hat{r}} =\mathbf{r}/r$. 
Similarly, in MG, a spinning particle/object creates a di-polar gravitomagnetic field $\mathbf{b}$ determined by its spin gravitomagnetic moment $\mathbf{\mu}_{gs}$ propertional to its spin angular momentum  $\mathbf{s}$. Using the classical methods of Kramers \cite{27}, Corben \cite{28} and Bohm \cite{29} for obtaining spin magnetic moment, in the case of MG, we found 
\begin{equation}
\mathbf{\mu}_{gs} =  \mathbf{s},  \label{eq7} 
\end{equation}
which is true for a uniformly gravito-magnetized sphere. It is to be noted that without doing detailed calculations as in refs. \cite{27,28,29}, one can quickly get the result in eq. \eqref{eq7} from eq. \eqref{eq5}, by replacing the electric charge $q$ with $m_0$, which represents the gravitational charge in the frame-work of MG as proved in ref. \cite{12,13}. Further, we note that for a spin $1/2$ Dirac particle Behera and Naik \cite{12} have obtained $\mathbf{\mu}_{gs} = \mathbf{s} = (\hbar/2)\mathbf{\sigma}$. The gravitational analogue of eq. \eqref{eq6}, in the case of MG is  
\begin{equation} 
\mathbf{b}(\mathbf{r}) = - \frac{\mu_{0g}}{4\pi}\left[\frac{3\mathbf{\hat{r}}(\mathbf{\hat{r} }\cdot \mathbf{\mu}_s) - \mathbf{\mu}_s }{r^3} \right] \label{eq8}  
\end{equation}
where $\mu_{0g} = 4\pi G/c^2$ is the gravito-magnetic permeability of vacuum. The ``minus" sign on the right hand side of eq.\eqref{eq8} is due to the fact that the source term ($\mathbf{j}_g$) in the grvito-Amp\'{e}re of MG, viz., 
$\mathbf{\nabla}\times \mathbf{b}  = - \mu_{0g}\mathbf{j}_g$ has a minus sign before it, in contrast with the positive sign before the corresponding term  ($\mathbf{j}_e$) in Amp\'{e}re's law ($\mathbf{\nabla}\times \mathbf{B}  = + \mu_{0}\mathbf{j}_e$) in electromagnetism. For the spinning Earth, $\mathbf{\mu}_{gs} = \mathbf{S}_E$, so the gravitomagnetic field of the Earth in the case of MG follows from eq. \eqref{eq8} as 
\begin{equation} 
\mathbf{b}(\mathbf{r}) = - \frac{G}{c^2}\left[\frac{3\mathbf{\hat{r}}(\mathbf{\hat{r} }\cdot \mathbf{S}_E) - \mathbf{S}_E}{r^3} \right].   \label{eq9}  
\end{equation}
\section{Spin Precession in Magnetic Field and Gravitomagnetic Field (Non-relativistic theory):}  
Let the spin magnetic moment of a particle with intrinsic spin $\mathbf{s}$, say an electron, be denoted by $\mathbf{\mu}_s = \alpha \mathbf{s}$, where $\alpha$ is proportionality constant (for electron $\alpha = q/m_0 = - |e|/m_0$). In non-relativistic classical theory of electromagnetism, if such a particle is placed in a magnetic field, its spin angular momentum $\mathbf{s}$ and hence spin axis changes with time as 
\begin{equation}
\frac{d\mathbf{s}}{dt} = \mathbf{\mu}_s \times \mathbf{B} = (-\alpha\mathbf{B}) \times \mathbf{s} = \mathbf{\Omega}_f \times \mathbf{s}  \label{eq10} 
\end{equation}
where $\mathbf{B}$ is the magnetic field at the position of the particle and the spin axis rotates with angular velocity 
\begin{equation}
\mathbf{\Omega}_f = -\alpha \mathbf{B}.  \label{eq11} 
\end{equation}
Similarly, in the non-relativistic (i.e., slow motion) formulation of gravito-electromagnetic (GEM) theory \cite{7,13} within domain of Newtonian physics, the gravito-magnetic analogues of eqs. \eqref{eq10}-\eqref{eq11} are 
\begin{eqnarray}
\frac{d\mathbf{s}}{dt} = \mathbf{\mu}_{gs} \times \mathbf{b}(0) = (-\mathbf{b}) \times \mathbf{s} = \mathbf{\Omega}_{fd}^{\text{GEM}} \times \mathbf{s} \label{eq12}  \\ 
\mathbf{\Omega}_{fd}^{\text{GEM}} = - \mathbf{b} = \frac{G}{c^2}\left[\frac{3\mathbf{\hat{r}}(\mathbf{\hat{r} }\cdot \mathbf{S}_E) - \mathbf{S}_E}{r^3} \right] = \mathbf{\Omega}_{fd}^{\text{GR}} \label{eq13}
\end{eqnarray}
which exactly matches with eq. \eqref{eq3} predicted by GR, if we consider $\mathbf{\mu}_{gs} = \mathbf{s}$ as the gravitomagnetic moment of a gyroscope in GP-B experiment and the gravitomagnetic field $\mathbf{b}$ of the Earth as given by eq. \eqref{eq9}. \\ 
\section{Spin Precession in Electric Field and Gravito-electric Field (Non-relativistic theory):} 
Suppose an electron with charge $q = -e$, rest mass $m_0$, spin magnetic moment $\mu_{es} = (q/m_0)\mathbf{s} = - (e/m_0) \mathbf{s}$ moves with velocity $\mathbf{v}$ in an external electric field $\mathbf{E}$. Then in the instantaneous rest frame of the electron it not only experiences the electric field $\mathbf{E}$ but also a magnetic field given by (in SI units)
\begin{equation}
\mathbf{B} = - \frac{\mathbf{v} \times \mathbf{E}}{c^2} = \frac{\mathbf{E} \times \mathbf{v}}{c^2}  \label{eq14}
\end{equation}
Then the equation of motion for its spin angular momentum $\mathbf{s}$ in its rest frame is 
\begin{equation}
 \left(\frac{d\mathbf{s}}{dt}\right)_{\text{rest}} = \mu_{es} \times  \mathbf{B} = - \left(\frac{q\mathbf{E} \times \mathbf{v}}{m_0c^2}\right) \times \mathbf{s} = \mathbf{\Omega}_{e}^I\times \mathbf{s}    \label{eq15}
\end{equation}
 and the spin axis of the electron precesses with an angular velocity 
 \begin{equation}
 \mathbf{\Omega}_{e}^I = - \left(\frac{q\mathbf{E} \times \mathbf{v}}{m_0c^2}\right) = \left(\frac{e\mathbf{E} \times \mathbf{v}}{m_0c^2}\right)  \label{eq16} 
 \end{equation}
In the case of an atomic electron moving around a nucleus of charge $Ze$ ($Z$ being the atomic number), the field $\mathbf{E} = (Ze \mathbf{r})/(4\pi \epsilon_0 r^3)$ and in this case eq. \eqref{eq16} becomes 
\begin{equation}
 \mathbf{\Omega}_{e}^I = \frac{Ze^2}{4\pi\epsilon_0 m_0 c^2 r^3  }\left(\mathbf{r}\times \mathbf{v}\right) \label{eq17} 
 \end{equation}
In the context of Maxwellain Gravity an analogous situation arises in the case of a spinning body moving in a Newtonian gravitoelectric field, just as a gyroscope moving around the Earth as in GP-B experiment. The quickest way to find the gravitational analogue of eq. \eqref{eq16}, is to replace $q\mathbf{E}$ by $m_0\mathbf{g} = - (G M m_0\mathbf{r})/r^3$, where $m_0$ now represents the mass of the gyroscope and $\mathbf{g}$ is the Newtonian gravito-electric field of the Earth of mass M; the result is 
\begin{equation}
\mathbf{\Omega}_g^I = - \mathbf{b} = \frac{GM}{c^2r^3}\left(\mathbf{r}\times \mathbf{v}\right) \label{eq18}
\end{equation} 
which is short of 
\begin{equation}
\mathbf{\Omega}_g^{II} = \frac{1}{2}\frac{GM}{c^2r^3}\left(\mathbf{r}\times \mathbf{v}\right) \label{eq19}
\end{equation} 
to get the correct GR value in eq. \eqref{eq2}. This deficiency is corrected from special relativistic contribution to the spin axis precession of an electron in its rest frame as calculated by Thomas \cite{31} in 1927, which we shall consider now. \\  
\section{Spin Precession in Electric Field and Gravito-electric Field (Relativistic Theory of Thomas):} For an electron with charge $q = - e$, rest mass $m_0$, spin angular momentum $\mathbf{s}$, spin magnetic moment $\mathbf{\mu}_{es}$, Thomas \cite{31} first calculated the relativistic change in the direction of its spin axis in electron's rest frame, which for the simple case of $\mathbf{\mu}_{es} = (q/m_0)\mathbf{s}$, is given by his eq. 4.122 in \cite{31} (written here in SI units and in different notations): 
\begin{eqnarray}
\frac{d\mathbf{s}}{d\tau} = \mathbf{\Omega}\times \mathbf{s}, \label{eq20} \\ 
\text{where}\, \quad \mathbf{\Omega} = - \frac{q}{m_0}\left(\mathbf{B}+\frac{1}{c^2}\frac{\gamma}{1+ \gamma}(\mathbf{E}\times \mathbf{v})\right),  \label{eq21}   
\end{eqnarray}
$\tau$ is the proper time, $\gamma= (1 - \mathbf{v}^2/c^2)^{-1/2}$ is the Lorentz factor, the fields $\mathbf{E}$ and $\mathbf{B}$ respectively refers to the electric and magnetic fields at the instantaneous position of the electron and $\mathbf{\Omega}$ is angular velocity of spin precession. For small velocity $|\mathbf{v}| << c$, $\gamma \rightarrow 1$ and the equation \eqref{eq21} reduces to 
  \begin{equation}
  \mathbf{\Omega} = - \frac{q}{m_0}\left(\mathbf{B}+\frac{1}{2c^2}(\mathbf{E}\times \mathbf{v})\right). \label{eq22}
  \end{equation}
If we put eq. \eqref{eq22} in eq. \eqref{eq20} we get Kramers equation (11) of 1934, in which Kramers \cite{27} followed a simpler approach. It should be clearly noted that the $1/2$ factor term in eq. \eqref{eq22} arises due to the relativistic $\gamma$ factor term in eq. \eqref{eq21}, which was absent in our non-relativistic approach in the previous section. Now substituting the value of $\mathbf{B}$ from eq. \eqref{eq14} in eq. \eqref{eq22}, we get 
\begin{equation}
\mathbf{\Omega}_{gd}^{Elect.} = - \frac{3}{2}\frac{q\mathbf{E}\times \mathbf{v}}{m_0c^2} = \frac{3}{2}\frac{Ze^2}{4\pi\epsilon_0 m_0 c^2 r^3  }\left(\mathbf{r}\times \mathbf{v}\right),    \label{eq23} 
\end{equation}
for an atomic electron, which is the electrical analogue of geodetic precession in eq. \eqref{eq2} in the atomic domain. \\
As was done in the previous section for the gravitational situation of GP-B experiment, we now replaced $q\mathbf{E}$ in eq. \eqref{eq23} by $m_0\mathbf{g} = - GMm_0\mathbf{r}/r^3$ to get the correct formula for geodetic precession according to HMG in flat space-time as
\begin{equation}
\mathbf{\Omega}_{gd}^{GEM} = - \frac{3}{2}\frac{m_0\mathbf{g}\times \mathbf{v}}{m_0c^2} = \frac{3GM}{2c^2r^3}\left(\mathbf{r}\times \mathbf{v}\right) = \mathbf{\Omega}_{gd}^{GR},    \label{eq24} 
\end{equation}
which is in perfect agreement with the prediction of GR in eq. \eqref{eq2}. One can also use equation (46) of Bailey \cite{32} to arrive at the eq.  \eqref{eq24} in the GEM frame work of HMG in flat space-time. With our results in eqs. \eqref{eq13} and \eqref{eq24}, we have thus shown that the GP-B experimental results may well be explained using Heaviside-Maxwellian Gravity (HMG) in flat space-time without the requirement of Einstein's General Relativity.
\section*{Conclusion}
The GP-B experimental data are in close agreement with the Schiff's formula for spin axis precession of a gyroscope in the gravitational field of the spinning Earth, which Schiff derived using Einstein's relativistic theory of gravity in curved space-time.  However, for the first time, here we report a Faraday-Maxwellian field theoretical explanation of the GP-B experimental results using Heaviside-Maxwellian (vector) Gravity (HMG) in flat space-time, first formulated by Heaviside in 1893, and later considered/re-discovered by many authors following a variant of approaches to gravito-electromagnetism. Our new explanation of the GP-B results involves an undergraduate level derivation the electrical and gravitational analogues of Schiff formula in flat space-time, which provides a new test of HMG apart from the existing ones noted in section 2. Well known physicist, C. M. Will concluded his viewpoint on GP-B results [Physics \textbf{4}, 43 (2011)] by stating, \textit{``The precession of a gyroscope in the gravitation field of a rotating body had never been measured before GP-B. While the results support Einstein, this didn't have to be the case. Physicists will never cease testing their basic theories, out of curiosity that new physics could exist beyond the ``accepted" picture.”}   



\begin{thebibliography}{}
\bibitem{1} Pugh G. E., WSEG research memorandum No. 11 (1959). 
\bibitem{2} Schiff, L.I. Possible new experimental test of general relativity theory. \href{https://doi.org/10.1103/PhysRevLett.4.215}{\textit{Phys. Rev. Lett}. \textbf{4}, 215 (1960).} 
\bibitem{3} Schiff, L.I. On experimental tests of the general theory of relativity. \href{https://doi.org/10.1119/1.1935800}{ \textit{Am. J. Phys}. \textbf{28}, 340 (1960).}
\bibitem{4} Schiff, L. I. Motion of gyroscope according to Einsteins theory of gravitation. \textit{ Proc. Natl. Acad. Sci. USA} \textbf{46}, 871 (1960).\href{https://www.pnas.org/content/pnas/46/6/871.full.pdf}{\textit{ Proc. Natl. Acad. Sci. USA} \textbf{46}, 871 (1960)}.
\bibitem{5} Everitt C. W. F., et al. Gravity Probe B: Final Results of a Space Experiment to Test General Relativity. \href{https://doi.org/10.1103/PhysRevLett.106.221101}{ \textit{Phys. Rev. Lett.} \textbf{106} 221101 (2011).}  
\bibitem{6} An online overview of the experiment is at \url{ http://einstein.stanford.edu}
\bibitem{7} Behera, H. Gravitomagnetism and Gravitational Waves in Galileo-Newtonian Physics. 2019). \href{https://arxiv.org/pdf/1907.09910.pdf}{arXiv:1907.09910}  
\bibitem{8} Behera, H., Barik, N. A New Set of Maxwell-Lorentz Equations and Rediscovery of Heaviside-Maxwellian (Vector) Gravity from Quantum Field Theory. (2019). \href{https://arxiv.org/pdf/1810.04791.pdf}{arXiv:1810.04791}
\bibitem{9} Heaviside, O. 1893. A Gravitational and Electromagnetic Analogy, Part I. \textit{The Electrician}, {\bf 31} 281-282 (1893). Part II. \textit{The Electrician}, {\bf 31} 359 (1893).
\bibitem{10} Heaviside O. \textit{Electromagnetic Theory}, vol.1, The Electrician Printing and Publishing Co., London, (1894). p. 455-465.  

\bibitem{11} O. Heaviside O. \textit{Electromagnetic Theory}, vol. 1, 3rd Ed. Chelsea Publishing Company, New York, (1971). p.455-466.
 
\bibitem{12} Behera, H., Naik, P. C. Gravitomagnetic Moments and Dynamics of Dirac (spin 1/2) Fermions in Flat Spape-time Maxwellian Gravity. \href{https://doi.org/10.1142/S0217751X04017768 }{ \textit{Int. J. Mod. Phys. A}, {\bf 19}, 4207-4229 (2004).} 

\bibitem{13} Behera, H. Comments on gravitoelectromagnetism of Ummarino and Gallerati in “Superconductor in a weak static gravitational field” vs other versions. \href{https://doi.org/10.1140/epjc/s10052-017-5386-4}{\textit {Eur. Phys. J. C.}, {\bf 77}: 822 (2017)}.

\bibitem{14} Abbott, B. P., et  al. Gravitational Waves and Gamma-Rays from a Binary Neutron Star Merger: GW170817 and GRB 170817A. \href{https://iopscience.iop.org/article/10.3847/2041-8213/aa920c}{\textit{The Astrophys. J. Lett.} {\bf 848} L13 (2017)}. 

\bibitem{15} Sattinger, D. H. Gravitation and Special Relativity. \href{https://doi.org/10.1007/s10884-013-9291-8}{ \textit{J. Dyn. Diff. Equat.}, {\bf 27} 1007-1025 (2015).}

\bibitem{16} Vieira, R. S., Brentan, H. B. 2018. Covariant theory of gravitation in the framework of special relativity. \href{https://doi.org/10.1140/epjp/i2018-11988-9} {\textit{Eur. Phys. J. Plus}, {\bf 133}, 165 (2018).}   

\bibitem{17} Jefimenko, O.  \textit{Gravitation and Cogravitation: Developing Newton's Theory of Gravitation to its Physical and Mathematical Conclusion.}, Electret Scientific Company, Star City, West Virginia,USA (2006).    

\bibitem{18} Jefimenko, O. \emph{Causality, electromagnetic induction, and gravitation : a different approach to the theory of electromagnetic and gravitational fields.}, 2nd Ed., Electret Scientific Company, Star City, West Virginia, USA (2000). 

\bibitem{19} Heras, J. A. An axiomatic approach to Maxwell's equations. \href{10.1088/0143-0807/37/5/055204}{\textit{Eur. J. Phys.} {\bf 37} 055204 (2016).} 
 
\bibitem{20} Nyambuya, G. G. Fundamental Physical Basis for Maxwell-Heaviside Gravitomagnetism. \href{ 10.4236/jmp.2015.69125}{\textit{ Journal of Modern Physics}, {\bf 6}, 1207-1219 (2015).} 
   
\bibitem{21} Ummarino G. A., Gallerati, A. Superconductor in a weak static gravitational field. \href{https://doi.org/10.1140/epjc/s10052-017-5116-y}{\textit{Eur. Phys. J. C.} {\bf 77}, 549 (2017).} 

\bibitem{22} Kennedy R. J. 1929. Planetary motion in a Retarded Newtonian Field. \href{https://doi.org/10.1073/pnas.15.9.744} {\textit{ Proc. N. A. S. \textbf {15}, 744.}} 
\bibitem{23} Singh A. 1982. Experimental Tests of the Linear Equations for the Gravitational Field. \href{https://doi.org/10.1007/BF02817094}{ \textit{ Lett. Nuovo Cimento}, \textbf {34}, 193-196}. 
\bibitem{24} Flanders W. D. Japaridze G. S. 2004. Photon deflection and precession of the periastron in terms of spatial gravitational fields. \href{https://iopscience.iop.org/article/10.1088/0264-9381/21/7/007}{\textit{Class. Quantum Grav. \textbf {21}, 1825-1831.}}
\bibitem{25} Borodikhin V. N. 2011. Vector Theory of Gravity. \href{https://link.springer.com/article/10.1134/S0202289311020071}{\textit{Gravitation and Cosmology, {\bf 17}, 161-165.}} 
\bibitem{26} Hilborn R. C. 2018. Gravitational waves from orbiting binaries without general relativity. \href{https://doi.org/10.1119/1.5020984}{\textit{  Am.J. Phys. \textbf {86}, 186.}} (2018). 

\bibitem{27} Kramers, H. A. On the classical Theory of the Spinning Electron. \href{https://www.sciencedirect.com/science/article/abs/pii/S0031891434802765}{\textit{ Physica}, \textbf{1}, 825-828(1934).}

\bibitem{28} Corben, H. C. 1993. Factors of 2 in magnetic moments, spin–orbit coupling, and Thomas precession. \href{http://dx.doi.org/10.1119/1.17207}{\textit{Am. J. Phys.} {\bf 61}, 551 (1993).}  

\bibitem{29} Bohm, A. \textit{Quantum Mechanics: Foundations and Applications, 3rd Ed.}. Springer-Verlag, New York (1993). p.262-264. 

\bibitem{30} Jackson J. D. \textit{Classical Electrodynamics, 3rd Ed.} John Wiley \& Sons (Asia) Pte. Ltd., Singapore (2004).

\bibitem{31} Thomas L. H. I. The kinematics of an electron with an axis, \href{http://dx.doi.org/10.1080/14786440108564170}{\textit{Phil. Mag. Ser.} \textbf{7, 3} :13, 1-22 (1927).} 

\bibitem{32} Bailey, Q. G. Lorentz-violating gravitoelectromagnetism. 
\href{https://journals.aps.org/prd/abstract/10.1103/PhysRevD.82.065012}{\textit{Phys. Rev. D}, \textbf{82}, 065012 (2010)}.





\end{thebibliography}
\end{document}